\documentclass[11pt,a4paper]{article}

\usepackage[utf8]{inputenc}
\usepackage[T1]{fontenc}
\usepackage{amsmath,amssymb,amsfonts}
\usepackage{graphicx}
\usepackage{booktabs}
\usepackage[hidelinks]{hyperref}
\usepackage{cleveref}
\usepackage{natbib}
\usepackage{algorithm}
\usepackage{algpseudocode}
\usepackage{listings}
\usepackage{xcolor}
\usepackage[margin=2.5cm]{geometry}

\definecolor{codegreen}{rgb}{0,0.6,0}
\definecolor{codegray}{rgb}{0.5,0.5,0.5}
\definecolor{codepurple}{rgb}{0.58,0,0.82}
\definecolor{backcolour}{rgb}{0.97,0.97,0.97}

\lstdefinestyle{pythonstyle}{
    backgroundcolor=\color{backcolour},
    commentstyle=\color{codegreen},
    keywordstyle=\color{blue},
    numberstyle=\tiny\color{codegray},
    stringstyle=\color{codepurple},
    basicstyle=\ttfamily\footnotesize,
    breakatwhitespace=false,
    breaklines=true,
    captionpos=b,
    keepspaces=true,
    numbers=left,
    numbersep=5pt,
    showspaces=false,
    showstringspaces=false,
    showtabs=false,
    tabsize=2,
    language=Python
}
\title{srvar-toolkit: A Python Implementation of Shadow-Rate Vector Autoregressions with Stochastic Volatility}

\author{
    Charles Shaw\thanks{Email: \href{mailto:charles@charlesshaw.net}{charles@charlesshaw.net}. ORCID: \href{https://orcid.org/0000-0002-3287-8251}{0000-0002-3287-8251}.}
}

\date{This version: 22 December 2025}

\begin{document}

\maketitle

\begin{abstract}
We introduce srvar-toolkit, an open-source Python package for Bayesian vector autoregression with shadow-rate constraints and stochastic volatility. The toolkit implements the methodology of \citet{grammatikopoulos2025forecasting} for forecasting macroeconomic variables when interest rates hit the effective lower bound. We provide conjugate Normal-Inverse-Wishart priors with Minnesota-style shrinkage, latent shadow-rate data augmentation via Gibbs sampling, diagonal stochastic volatility using the Kim-Shephard-Chib mixture approximation, and stochastic search variable selection. Core dependencies are NumPy, SciPy, and Pandas, with optional extras for plotting and a configuration-driven command-line interface. We release the software under the MIT licence at \url{https://github.com/shawcharles/srvar-toolkit}.

\end{abstract}

\section{Introduction}
\label{sec:introduction}

Vector autoregressions have anchored macroeconomic forecasting since \citet{sims1980macroeconomics} demonstrated their power for capturing dynamic relationships among economic variables. Yet when central banks push policy rates to zero---the effective lower bound (ELB)---the VAR framework faces a challenge. The observed rate no longer reflects the true stance of monetary policy. Shadow-rate models resolve this tension by positing a latent rate that can turn negative, with the observed rate censored at the bound \citep{wu2016measuring}. This paper introduces \texttt{srvar-toolkit}, a Python implementation that makes shadow-rate VAR estimation accessible to researchers who prefer open-source tools.

\citet{grammatikopoulos2025forecasting} develops the statistical framework we implement here, combining shadow-rate data augmentation with Bayesian shrinkage priors and stochastic volatility. The original MATLAB code serves the econometrics community well, but many applied researchers now work primarily in Python. We bridge that gap. Our toolkit offers a transparent, tested implementation with a clean separation between model specification, prior configuration, and MCMC sampling. Users can fit a basic Bayesian VAR in a few lines of code, then progressively add complexity---Minnesota shrinkage, ELB constraints, time-varying volatility---without changing the core workflow.

The architecture reflects four design commitments. We use immutable configuration objects to prevent subtle bugs from mutable state. We inject random number generators explicitly to ensure reproducibility. We limit core dependencies to NumPy, SciPy, and Pandas, avoiding the complexity of probabilistic programming frameworks; plotting and the configuration-driven CLI are provided via optional extras. We maintain a test suite that pins down numerical behaviour across releases. The result is a toolkit suited to both exploratory analysis and production forecasting pipelines.

\Cref{sec:methodology} summarises the statistical methodology. \Cref{sec:software} describes the software architecture. \Cref{sec:demonstration} presents an empirical illustration. \Cref{sec:conclusion} discusses limitations and future directions.

\section{Methodology}
\label{sec:methodology}

This section sketches the statistical models we implement. For derivations and theoretical foundations, see \citet{grammatikopoulos2025forecasting}.

\subsection{The VAR Model and Conjugate Priors}

Consider $N$ macroeconomic variables observed over $T$ periods. The VAR($p$) model expresses each variable as a linear function of its own lags and lags of all other variables in the system. We write the model as
\begin{equation}
    y_t = c + B_1 y_{t-1} + \cdots + B_p y_{t-p} + \varepsilon_t, \quad \varepsilon_t \sim \mathcal{N}(0, \Sigma),
\end{equation}
where $y_t$ collects the $N$ variables, $c$ is an intercept vector, and $B_1, \ldots, B_p$ are coefficient matrices. The error covariance $\Sigma$ captures contemporaneous relationships. Stacking observations into matrices $Y$ (responses) and $X$ (lagged values plus intercept), we obtain the compact form $Y = XB + E$ with $B$ a $K \times N$ coefficient matrix and $K = 1 + Np$.

Bayesian estimation requires priors on $(B, \Sigma)$. The conjugate choice is Normal-Inverse-Wishart, where
\begin{equation}
    B \mid \Sigma \sim \mathcal{MN}(M_0, V_0, \Sigma) \quad \text{and} \quad \Sigma \sim \mathcal{IW}(\nu_0, S_0).
\end{equation}
The matrix-normal distribution $\mathcal{MN}$ places a prior mean $M_0$ on the coefficients with row precision governed by $V_0$ and column covariance inherited from $\Sigma$. The inverse-Wishart prior on $\Sigma$ has degrees of freedom $\nu_0$ and scale matrix $S_0$. Conjugacy means the posterior takes the same NIW form with updated parameters $(M_n, V_n, S_n, \nu_n)$, enabling fast closed-form sampling.

\subsection{Minnesota Shrinkage}

Unrestricted VARs with many variables and lags suffer from proliferating parameters. Minnesota-style shrinkage \citep{doan1984forecasting} addresses this by tightening the prior around parsimonious specifications. We construct $V_0$ so that own lags receive more prior weight than cross-variable lags, and recent lags receive more weight than distant ones. The key hyperparameter $\lambda_1$ controls overall tightness, while $\lambda_3$ governs lag decay. Larger values of $\lambda_1$ pull coefficients towards the prior mean; smaller values let the data speak. We estimate residual variances from univariate AR regressions to scale the prior appropriately across variables with different units.

\subsection{Shadow-Rate Data Augmentation}

When a policy rate hits the ELB, the observed value $y_{t,j}^{\text{obs}}$ no longer equals the latent ``shadow rate'' $y_{t,j}^*$ that would prevail without the constraint. We model the censoring as
\begin{equation}
    y_{t,j}^{\text{obs}} = \max(y_{t,j}^*, b),
\end{equation}
where $b$ is the bound. At periods when the observed rate lies at or near the bound, we treat the shadow rate as a latent variable and sample it from its conditional distribution. Given the VAR parameters and the other variables, the shadow rate follows a truncated normal distribution on $(-\infty, b]$. We draw from this distribution at each MCMC iteration, alternating with draws of the VAR coefficients. This data augmentation approach, detailed in \citet{grammatikopoulos2025forecasting}, respects the censoring while propagating uncertainty about the latent values into forecasts.

\subsection{Stochastic Volatility}

Macroeconomic volatility changes over time. The Great Moderation saw declining variance; the 2008 crisis brought a spike. A constant $\Sigma$ misses this variation. We adopt diagonal stochastic volatility, where each variable has its own time-varying log-variance following a random walk:
\begin{equation}
    \varepsilon_{t,i} = \exp(h_{t,i}/2) \, \eta_{t,i}, \quad h_{t,i} = h_{t-1,i} + \xi_{t,i},
\end{equation}
with $\eta_{t,i} \sim \mathcal{N}(0,1)$ and $\xi_{t,i} \sim \mathcal{N}(0, \sigma_{\eta,i}^2)$. The log-volatility $h_{t,i}$ evolves smoothly, capturing persistent shifts in uncertainty.

Estimation poses a challenge because the observation equation is nonlinear in $h_{t,i}$. \citet{kim1998stochastic} show that squaring and taking logs transforms the problem into a linear state-space model contaminated by $\log \chi^2_1$ errors. They approximate this non-Gaussian error with a seven-component mixture of normals, enabling standard Kalman filtering and smoothing. We implement their precision-based sampler, which exploits the banded structure of the state-space precision matrix for efficient draws.

\subsection{Variable Selection}

With many potential predictors, some coefficients may be effectively zero. Stochastic search variable selection \citep{george1993variable} formalises this intuition through spike-and-slab priors. In our implementation, each predictor row of the VAR coefficient matrix (including lagged regressors and, optionally, the intercept) has an inclusion indicator that selects either a tight ``spike'' variance or a diffuse ``slab'' variance shared across equations. We sample these row-wise inclusion indicators within the Gibbs sampler, letting the data determine which predictors matter. The intercept can be forced to remain included to avoid shrinking the unconditional mean.

\section{Software Design}
\label{sec:software}

\subsection{Architecture}

We organise \texttt{srvar-toolkit} into modules that separate concerns cleanly. The \texttt{srvar.api} module exposes the two functions most users need: \texttt{fit()} runs the MCMC sampler and returns posterior draws, while \texttt{forecast()} generates predictive simulations from a fitted model. Configuration lives in \texttt{srvar.spec}, where dataclasses define model structure (\texttt{ModelSpec}), prior parameters (\texttt{PriorSpec}), and sampler settings (\texttt{SamplerConfig}). The statistical workhorses reside in specialised modules: \texttt{srvar.bvar} handles NIW posteriors, \texttt{srvar.elb} implements shadow-rate sampling, \texttt{srvar.sv} provides the KSC mixture sampler for stochastic volatility, and \texttt{srvar.ssvs} manages spike-and-slab variable selection. Data ingestion and transformation utilities live in \texttt{srvar.data}, while \texttt{srvar.plotting} offers visualisations for shadow rates, volatility paths, and forecast fan charts.

This modular structure lets users replace or extend components without touching unrelated code. A researcher who wants a different volatility specification can subclass or replace \texttt{srvar.sv} while keeping the rest of the pipeline intact.

\subsection{Design Principles}

Four principles guide our implementation. First, we make configuration objects immutable. Once you create a \texttt{ModelSpec}, you cannot accidentally mutate it mid-estimation. This prevents a class of subtle bugs where shared mutable state leads to irreproducible results. Second, we inject random number generators explicitly. Every sampling function accepts an \texttt{rng} argument of type \texttt{numpy.random.Generator}. Callers control the seed; the library never touches global state. Third, we limit core dependencies to NumPy, SciPy, and Pandas, avoiding the complexity of probabilistic programming frameworks; plotting and the configuration-driven CLI are provided via optional extras. Fourth, we test aggressively. The test suite exercises core samplers with fixed seeds, checking that outputs match expected values across releases and platforms.

\subsection{Usage Example}

A minimal workflow loads data, configures the model, and runs estimation. The listing below fits a four-lag VAR with Minnesota shrinkage to a three-variable system:

\begin{lstlisting}[caption={Fitting a Bayesian VAR with Minnesota shrinkage.}]
import numpy as np
from srvar import Dataset
from srvar.api import fit, forecast
from srvar.spec import ModelSpec, PriorSpec, SamplerConfig

ds = Dataset.from_arrays(
    values=data_array,
    variables=["gdp", "inflation", "rate"]
)

model = ModelSpec(p=4, include_intercept=True)
prior = PriorSpec.niw_minnesota(p=4, y=ds.values, n=ds.N)
sampler = SamplerConfig(draws=2000, burn_in=500, thin=2)

result = fit(ds, model, prior, sampler, rng=np.random.default_rng(42))
fc = forecast(result, horizons=[1, 4, 8], draws=500)
\end{lstlisting}

Adding shadow-rate constraints and stochastic volatility requires only minor changes to the model specification:

\begin{lstlisting}[caption={Enabling ELB constraints and stochastic volatility.}]
from srvar import ElbSpec, VolatilitySpec

model = ModelSpec(
    p=4,
    include_intercept=True,
    elb=ElbSpec(applies_to=["rate"], bound=0.125),
    volatility=VolatilitySpec(enabled=True)
)

result = fit(ds, model, prior, sampler, rng=np.random.default_rng(42))
\end{lstlisting}

The \texttt{ElbSpec} names the constrained variable and sets the bound; \texttt{VolatilitySpec} activates time-varying variances. The same \texttt{fit()} and \texttt{forecast()} calls handle the expanded model.

\section{Demonstration}
\label{sec:demonstration}

We illustrate the toolkit with a small, fully reproducible example consistent with the repository workflows. The demo uses a two-variable monthly system in which an observed short rate is censored at an ELB and a second macro variable responds to lagged rates. This setup mirrors the shadow-rate mechanism while remaining lightweight enough to run quickly on a laptop.

\begin{figure}[htbp]
    \centering
     \includegraphics[width=0.8\textwidth]{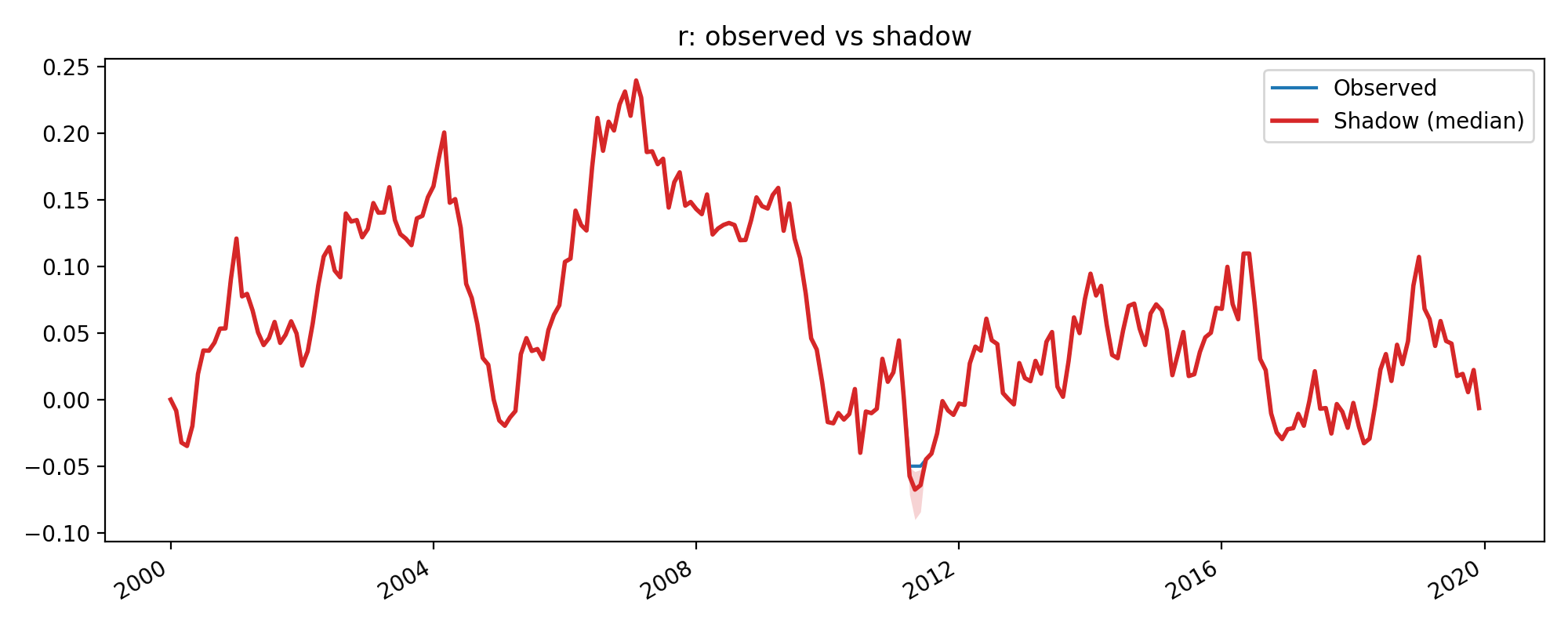}
    \caption{Inferred shadow rate (solid) versus observed rate (dashed) for the ELB-censored series.}
    \label{fig:shadow_rate}
\end{figure}

\Cref{fig:shadow_rate} plots the posterior median shadow rate alongside the observed censored rate. When the ELB binds, the inferred shadow rate can fall below the bound, capturing accommodation that is not visible in the observed series.

\begin{figure}[htbp]
    \centering
     \includegraphics[width=0.8\textwidth]{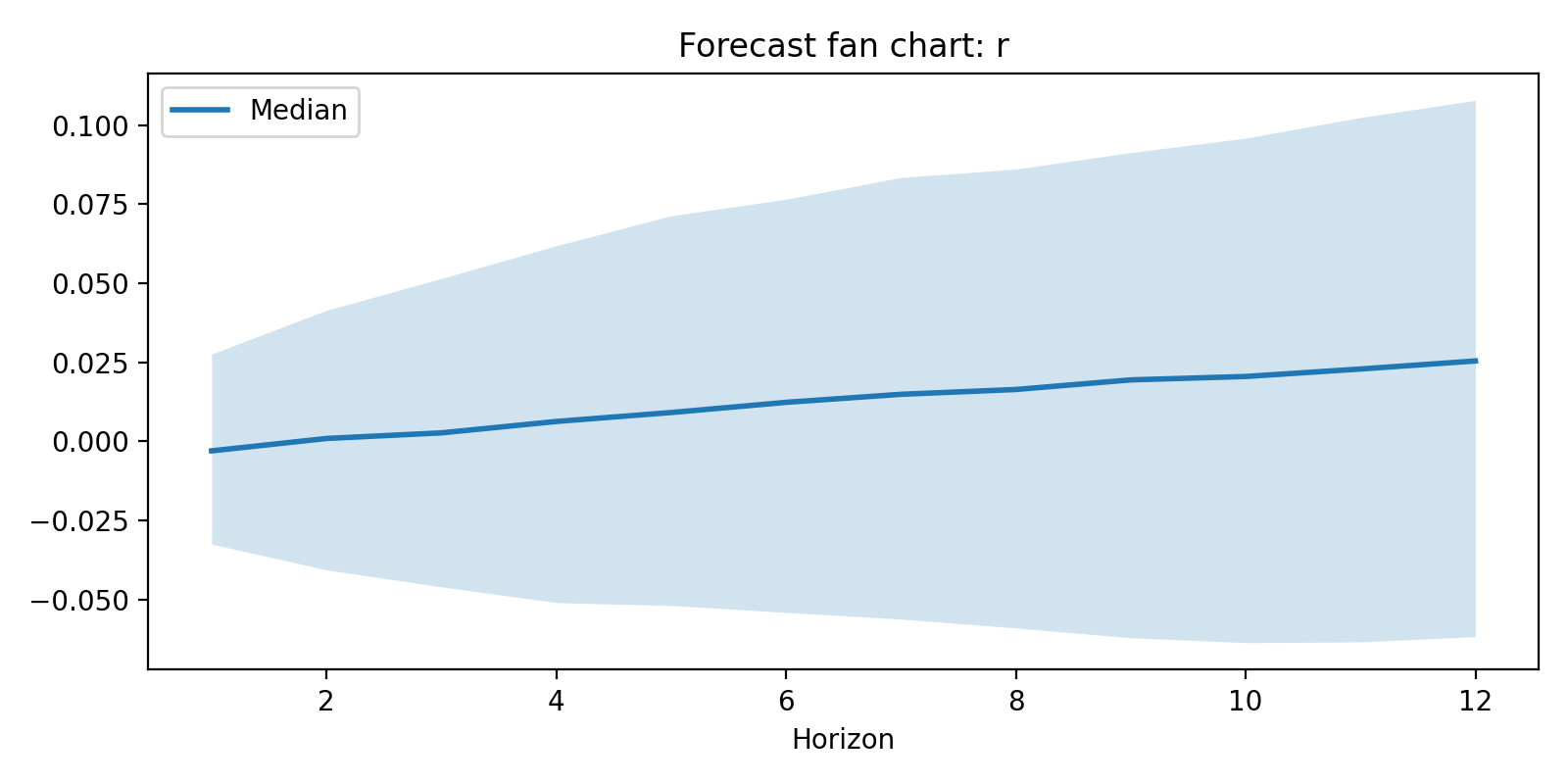}
    \caption{Forecast fan chart for the shadow-rate series showing 10\%, 50\%, and 90\% quantiles.}
    \label{fig:forecast}
\end{figure}

\Cref{fig:forecast} shows forecast fan charts generated from the combined ELB-SV model. The widening bands reflect both parameter uncertainty and time-varying volatility.

\section{Conclusion}
\label{sec:conclusion}

We have introduced \texttt{srvar-toolkit}, a Python package for Bayesian VAR estimation with shadow-rate constraints and stochastic volatility. The software implements the methodology of \citet{grammatikopoulos2025forecasting} in a transparent, tested framework that separates model specification from sampling machinery.

The current implementation has limitations worth noting. Stochastic volatility is diagonal, so we do not model time-varying correlations between variables. Steady-state VAR parameterisation, useful for imposing long-run equilibrium conditions, remains unimplemented. We plan to add alternative shrinkage priors---Bayesian LASSO, Dirichlet-Laplace---in future releases. Despite these gaps, the toolkit covers the core workflow needed for applied shadow-rate VAR analysis.

The software is actively maintained. Documentation, including installation instructions and worked examples, is available at \url{https://srvar-toolkit.charlesshaw.net/}. We welcome contributions and bug reports through the GitHub repository.

\bibliographystyle{apalike}
\bibliography{references}

\end{document}